\documentclass[preprint, 
amsmath,amssymb, aps, 12pt, longbibliography]{revtex4-1}

\usepackage{amssymb}
\usepackage{amsmath}
\usepackage{bm}
\usepackage{graphics}
\usepackage{graphicx}
\usepackage{soul}
\usepackage{color}
\usepackage{placeins} %uso el floatbarrier para separar bibliografía
\usepackage{dblfloatfix}
\usepackage{textcomp}
\usepackage{dcolumn}% Align table columns on decimal point
\usepackage[font={footnotesize}]{caption}
\begin{document}

%\begin{frontmatter}

\title{Vortex dynamics under pulsatile flow in axisymmetric constricted tubes
}


\author{Nicasio Barrere, Javier Brum, Alexandre L'Her, Gustavo L. Saras\'ua, Cecilia Cabeza}

\affiliation{Instituto de F\'isca, Facultad de Ciencias, UdelaR, Uruguay}
%\address{nbarrere@fisica.edu.uy}

\begin{abstract}
An improved understanding of how vortices develop and propagate under pulsatile flow can shed important light on the mixing and transport processes including the transition to turbulent regime occurring in such systems. For example, the characterization of pulsatile flows in obstructed artery models serves to encourage research into flow-induced phenomena associated with changes in morphology, blood viscosity, wall elasticity and flow rate. In this work, an axisymmetric rigid model was used to study the behaviour of the flow pattern with varying constriction degree ($d_0$), mean Reynolds number ($\bar{Re}$) and Womersley number ($\alpha$). Velocity fields were acquired experimentally using Digital Particle Image Velocimetry and generated numerically. For the acquisition of data, $\bar{Re}$ was varied from 385 to 2044, $d_0$ was 1.0 cm and 1.6 cm, and  $\alpha$ was varied from 17 to 33 in the experiments and from 24 to 50 in the numerical simulations. Results for the considered Reynolds number, showed that the flow pattern consisted of two main structures: a central jet around the tube axis and a recirculation zone adjacent to the inner wall of the tube, where vortices shed. Using the vorticity fields, the trajectory of vortices was tracked and their displacement over their lifetime calculated. The analysis led to a scaling law equation for the maximum vortex displacement as a function of a dimensionless variable dependent on the system parameters Re and $\alpha$. 

\begin{description}
\item[PACS]
{47.20.Ib Instability of boundary layers; separation},       {47.32.cb	Vortex interactions}
\end{description}

\end{abstract}


%\end{frontmatter}
\maketitle

\section{Introduction}
Research on the dynamics of pulsatile flows through constricted regions has multiple applications in biomedical engineering and medicine. Cardiovascular diseases are the primary cause of death worldwide (accounting for 31\% of total deaths in 2012) \cite{naghavi}. More than half of these deaths could have been avoided by prevention and early diagnosis. Atherosclerosis is characterized by the accumulation of fat, cholesterol and other substances in the intima layer, creating plaques which obstruct the arterial lumen. Atherosclerotic plaque fissuring and/or breaking are the major causes of cardiovascular stroke and myocardial infarct. Several factors leading to plaque complications have been reported: sudden increase in luminal pressure \cite{muller}, turbulent fluctuations \cite{loree}, hemodynamic shear stress \cite{gertz}, vasa vasorum rupture \cite{barger}, material fatigue \cite{falk} and stress concentration within a plaque \cite{imoto}. As most of these factors are flow-dependent, understanding how a pulsatile flow behaves through a narrowed region can provide important insight toward the development of reliable diagnostic tools.

Flow alterations due to arterial obstruction (i.e. stenosis) and aneurysms have been reported in the literature \cite{caro,ku_1985,nerem_87,Gopalakrishnan2014}. Simplified models of stenosed arteries have used constricted rigid tubes \cite{arzani,ford,boussel,long_2001}. At moderate Reynolds numbers, the altered flow splits at the downstream edge of the constriction into a high-velocity jet along the centreline and a vortex shedding zone separating from the inner surface of the tube wall (recirculation flow). Increasing Reynolds numbers lead to a transition pattern and ultimately turbulence. 

Several experimental studies have been reported. Stettler and Hussain \cite{sadan} studied the transition to turbulence of a pulsatile flow in a rigid tube using one-point anemometry. However, one-point anemometry fails to account for different flow structures which may play an important role at the onset of turbulence. Peacock \cite{peacock} studied the transition to turbulence in  a  rigid  tube  using  two-dimensional  particle  imaging  velocimetry  (PIV). Chua et al \cite{chua_piv_2009} implement a three-dimensional PIV technique to get a volumetrical velocity field of a steady flow.
The work of Ahmed and Giddens \cite{ahmed_83, ahmed_84} studied the transition to turbulence in a constricted rigid tube with varying degree of constriction using two-component laser Doppler anemometry. For constriction degrees up to 50\% of the lumen, the authors found no disturbances at Reynolds numbers below 1000. This study was not focused on describing vortex dynamics, however, the authors reported that turbulence, when observed, was preceded by vortex shedding. 

Several authors carried out numerical studies.  Long et al. \cite{long_2001} compared the flow patterns produced by an axisymmetric and a non-axisymmetric geometry finding that flow instabilities through an axisymmetric geometry was more sensitive to changes in the degree of constriction. The work of Isler et al. \cite{Isler2018} in  an axisymmetric constricted channel found the instabilities that breaks the symmetry  of the flow. The works of Mittal et al. \cite{mittal} and Sherwin and Blackburn \cite{sherwin_2005} studied the transition to turbulence over a wide range of Reynolds numbers. Mittal et al. used a planar channel with a one-sided semicircular constriction. They found that downstream of the constriction, the flow was composed of two shear layers, one originating at the downstream edge of the constriction and another separating from the opposite wall. For Reynolds numbers above 1000, the authors reported transition to turbulence due to vortex shedding. Moreover, they found through spectral analysis that the characteristic shear layer frequency is associated with the frequency of vortex formation. Sherwin and Blackburn \cite{sherwin_2005} studied the transition to turbulence using a three-dimensional axisymmetric geometry with a sinusoidal constriction. Based on the results of linear stability analysis, the authors reported the occurrence of Kelvin-Helmholtz instability. This reaffirms that instabilities involving vortex shedding take place in the transition to turbulence.

Finally, several studies compared experimental and numerical results. Ling et al. \cite{ling_72} compared numerical results with those obtained by hot-wire measurements. As mentioned above, one-dimensional hot-wire measurements cannot be used to identify flow structures. Griffith et al. \cite{griffith}, using a rigid tube with a slightly narrowed section resembling stenosis, compared numerical results with experimental data. Using stability analysis by means of Floquet exponents, they demonstrated that the experimental flow was less stable than that of the simulated model. For low Reynolds numbers (50 $-$ 700), the authors found that a ring of vortices formed immediately downstream of the stenosis and that its propagation velocity changed with the degree of stenosis.  The work of Usmani and Muralidhar \cite{usmani} compares flow patterns in rigid and compliant asymmetric constricted tubes for Reynolds range between 300 and 800 and Womersley between 6 and 8. The authors report that the downstream flow is characterized by a high velocity jet and a vortex whose evolution is qualitatively described. The aforementioned studies highlight the significance of study the vortex dynamics since vortex development precedes turbulence and ultimately, contributes to the risk of a cardiovascular stroke. 

The aim of this work is to characterize the vortex dynamics under pulsatile flow in an axisymmetric constricted rigid tube. The study was carried out both experimentally and numerically for different constriction degrees and varying mean Reynolds number from 385 to 2044 and Womersley number from 17 to 50. %covering  a range of applications from human arteries ($\alpha$=17.8 for ascending aorta \cite{san_2012}) to  typical values in engineering applications \cite{high_freq_pulsatile_2011}.
These results show that the flow pattern in these systems consists primarily of a central jet and a vortex shedding layer adjacent to the wall (recirculation flow), consistent with the literature. By tracking the vortex trajectory, it was possible to determine the displacement of vortices over their lifetime. The vortex kinematics was described as a function of the system parameters in the form of a dimensionless scaling law for the maximum vortex displacement.

\section{Methods and materials}
\subsection{Experimental setup}
\begin{figure}[h!]
 \resizebox{1\textwidth}{!}{%
   \includegraphics{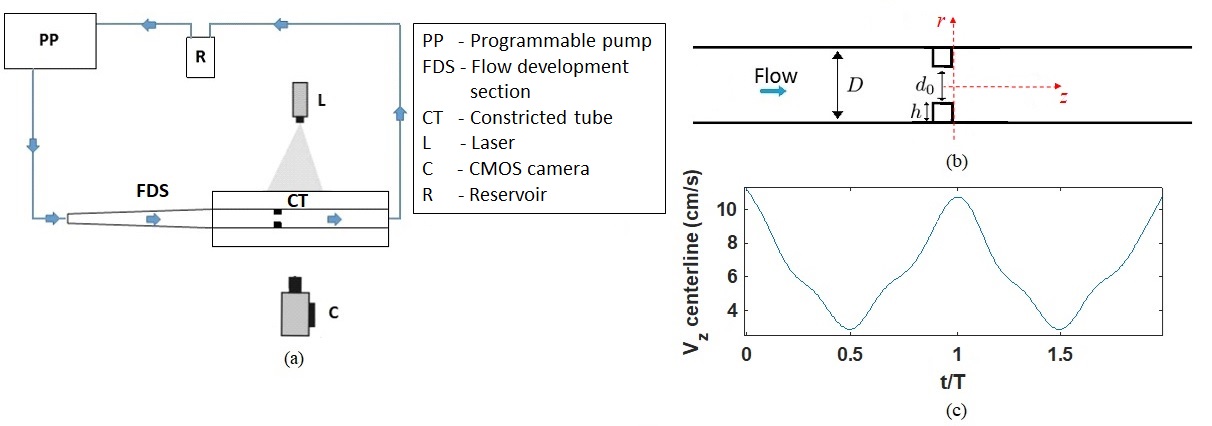}}
   \caption{: (a) Schematic drawing of the experimental setup; (b) lengthwise view of a tube of diameter $D = 2.6 \pm 0.1$ cm and an annular constriction of diameter $d_0$; (c) experimental axial velocity at $z = 0$ and $r = 0$ for $\Bar{Re}$ = 1002, with $d_0$ = 1.6 cm .}
 \label{fig:setup}       % Give a unique label
 \end{figure}

The experimental setup (Fig. \ref{fig:setup} (a)) consisted of a circulating loop and a Digital Particle Imaging Velocimetry (DPIV) module capable of acquiring videos and processing velocity fields.
The circulating loop consisted of a programmable pulsatile pump (PP), a reservoir (R), a flow development section (FDS) and a tube containing an annular constriction (CT). 

The tube with the annular constriction (CT) consisted of a transparent acrylic rigid tube with $L=51.0\pm 0.1$ cm in length and inner diameter $D = 2.6 \pm 0.1$ cm. The annular constriction consisted of a hollow cylinder $5\pm 1$ mm in axial length, which was fitted inside the rigid tube. Keeping the outer diameter of the annular constriction fixed at 2.6 cm, i.e., equal to the inner diameter of the rigid tube, its inner diameter was changed to achieve different degrees of constriction. Tests were carried out using annular constrictions with inner diameters $d_0 = 1.6 \pm 0.1$ cm and $d_0 = 1.0 \pm 0.1$ cm, equivalent to degrees of constriction relative to $D$ of 39\% and 61\%, respectively. A cross-sectional scheme of the constriction geometry is shown in Fig. \ref{fig:setup} (b), along with the coordinate system used throughout the study.  The radial direction is represented by $r$ and the axial direction by $z$, with $r$ = 0 coinciding with the tube axis and $z$=0 with the downstream edge of the constriction. Within this axis representation, the downstream region is defined by $z>$0 values and the upstream region is defined by $z<0$ values. Finally, this constricted tube (CT) was placed inside a chamber filled with water, so that the refraction index of the fluid inside the tube matched that of the outside fluid.

 The tube inlet was connected to the pulsatile pump via a flow development section (FDS), whereas its outlet was connected to the reservoir. The flow development section was designed to ensure a fully developed flow at the tube inlet and consisted of two sections: first, a conical tube of 35 cm in length with inner diameter increasing from 1 cm to 2.6 cm to provide a smooth transition from the outlet of the pump. Secondly, an acrylic rigid tube of inner diameter $D$ and length of 48$D$ connected to the constricted tube (CT). The reservoir (R) was used to set the minimum pressure inside the tube, which was set equal to atmospheric pressure in the experiments.
 
 The system was filled with degassed water and seeded with neutrally buoyant polyamide particles (0.13 g/l concentration, 50 $\mu$m diameter, DANTEC). DPIV technique was used to acquire the velocity field  \cite{piv1,piv2}. A 1 W Nd:YAG laser was used to illuminate a 2 mm-thick section of the tube. Images were acquired at a frame rate of 180 Hz over a period equivalent to 16 cycles, using a CMOS camera (Pixelink, PL-B776F). The velocity field was finally computed using OpenPIV open-source software with $32 \times 32$ pixel$^2$ windows and an overlap of 8 pixels in both directions. 
 
 The region of interest was defined as -0.5$<r/D<$0.5, 0$< z/D <$1.5. Within this region no turbulence was observed, as confirmed by spectral analysis of the velocity fields. This observation is also consistent with previous observations \cite{peacock,ling_72, casanova}. Due to the pulsatility and the absence of turbulence, it was possible to consider each cycle as an independent experiment, using the ensemble average over all 16 cycles to obtain the final velocity fields.

For each degree of constriction, different experiments were carried out varying the flow velocity at the tube inlet. The pulsation period and the shape of the velocity as function of $t$, in $z$ = 0 and $r$ = 0, shown in Fig.\ref{fig:setup} (c), remained unchanged throughout the experiments. Pulsation period and peak velocity can be related to the Womersley and Reynolds numbers, respectively. The Womersley number, $\alpha$, is defined as $\alpha=\frac{D}{2}\sqrt{\frac{2 \pi f}{\nu}}$, where $f$ is the pulsatile frequency and $\nu$ the kinematic viscosity of water ($\nu = 1.0 \times 10^{-6}$ m$^2$/s).

The experiments, were carried out for three different pulsatile period $T$: 0.96s, 2.39s and 3.58s corresponding to $\alpha$ values of 33.26, 21.10 and 17.22 respectively. The peak Reynolds number upstream of the constriction is defined as $Re_\mathrm{u} = D v_\mathrm{u}/\nu$ where $v_\mathrm{u}$ is the peak velocity measured at the centreline. Four different values of $Re_\mathrm{u}$ were used as inlet condition upstream of the constriction, see table \ref{tab:1}. Similarly, the mean Reynolds at the downstream edge of the constriction is defined as $\Bar{Re} = d_0 \Bar{v}/\nu$,  where $\Bar{v}$ is the mean velocity over a whole period, measured at $(z,r)=(0,0)$.  Finally, experiments are labeled through its $\bar{Re}$ values since it represents a unique combination of $Re_\mathrm{u}$, $d_0$ and $\alpha$ as summarized in table \ref{tab:1}. 

\begin{table}[h!]
\centering
\begin{tabular}{ccccc}

\hline\noalign{\smallskip}
 $\bm{\alpha}$ & $\bm{\bar{Re}}$ & $Re_\mathrm{u}$ &Constriction & $d_0$  (cm)  \\
\noalign{\smallskip}\hline\noalign{\smallskip}
 \textbf{33.26} & \textbf{654} & 820 &39\% &  1.6\\
                &\textbf{1002}& 820 &61\% & 1.0 \\
                &\textbf{1106}& 1187 &39\% &  1.6\\
                 %12001 & 1200 & 1950 &39\% \\
                &\textbf{1767}& 1187 &61\% & 1.0 \\
                &\textbf{2044}& 1625 &61\% & 1.0 \\
   
%\noalign{\smallskip}
\hline    
\textbf{21.10} & \textbf{385}& 505  &39\% & 1.6 \\ %634
               & \textbf{963}& 820 &39\% & 1.6 \\%1496
\hline 
\textbf{17.22} & \textbf{585}& 820 &39\% & 1.6 \\%1198
              % & \textbf{895}& 1187 &39\% & 1.6 \\%1419               
\end{tabular}
\caption{Experimental degrees of constriction and $\Bar{Re}$  numbers.
}
\label{tab:1} 
\end{table}

\begin{figure}[h!]
\resizebox{0.5\textwidth}{!}{%
  \includegraphics{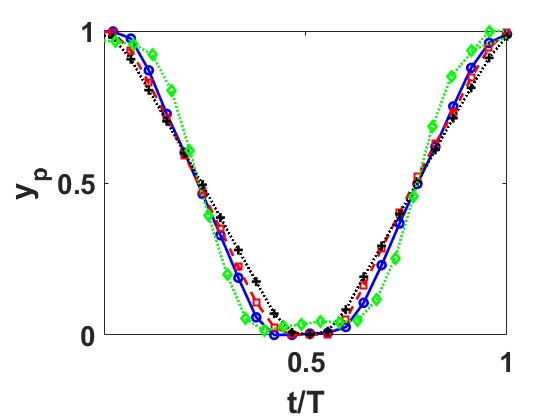}}
  \centering
\caption{Normalized displacement of piston $y_p$ for $Re_\mathrm{u}$=505 (diamond green), $Re_\mathrm{u}$=820 (blue circles), $Re_\mathrm{u}$=1187 (red squares), $Re_\mathrm{u}$=1625 (black plus sign).}
\label{fig:piston}       % Give a unique label
\end{figure}

The pulsatile pump (PP in Fig. \ref{fig:setup} (a)) generated different flow conditions at the tube inlet, \cite{balay}. The pumping module consisted of a step-by-step motor driving a piston inside a cylindrical chamber. The control module contained the power supply for all the system and an electronic board, programmable from a remote PC by means of a custom software. The piston motion was programmed using the pump settings for pulsatile period, ejected volume, pressure and cycle shape. The piston motion of the programmable-pump is showed in fig. \ref{fig:piston}  in normalized units of displacement and time. Figure \ref{fig:piston} shows that the piston motion, and hence the pulsatile waveform  are comparable among all $Re_\mathrm{u}$ values, meaning that the inlet conditions are equivalent for all the experiments. 

\begin{figure}[h!]
\resizebox{0.87\textwidth}{!}{%
  \includegraphics{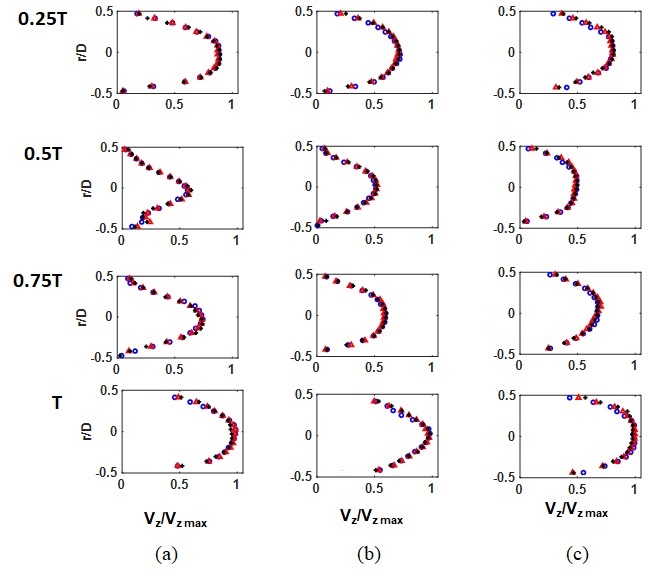}}
  \centering
\caption{Upstream velocity profiles for (a) $Re_\mathrm{u}$=820, (b) $Re_\mathrm{u}$=1187  and (c)  $Re_\mathrm{u}$=1625. Profiles are located at $z_1$=-1.25D (blue circles), $z_2$=-0.75D (red triangles) and $z_3$=-0.25D (black plus sign).}
\label{fig:profiles}       % Give a unique label
\end{figure}

Finally, the condition of fully developed flow at the upstream region was verified.   The entrance length is usually estimated as $L\approx 0.05ReD$.   According to the work of Ku \cite{ku_1997}, the entrance length of a pulsatile flow with $\alpha>$10, corresponds to the upstream mean Reynolds over a whole cycle, $\Bar{Re}_\mathrm{u}$, giving  $L\approx 0.05\Bar{Re}_\mathrm{u}D$. In our setup, the  flow development section measures 48D, $\Bar{Re}_\mathrm{u}<$1000 and $\alpha>$10 for all experiments, which makes reasonable assuming developed flow. Nevertheless, velocity profiles were studied for  $Re_\mathrm{u}$ values of 820, 1187 and 1625 in order to ensure developed flow condition.  Profiles are taken from a $1.5D$ length window upstream of the constriction, at fixed locations $z_1=-1.25D$, $z_2=-0.75D$ and $z_3=-0.25D$. Velocity values are normalized by the maximum velocity attained in a pulsatile cycle. Figure \ref{fig:profiles} shows such profiles at $t=0.25T$, $t=0.5T$, $t=0.75T$ and $t=T$. Profiles shows that for each $Re_\mathrm{u}$ independently of the time, profiles are equal for $z_1$=-1.25D, $z_2$=-0.75D and $z_3$=-0.25D which demonstrates that the flow is developed.

\subsection{Numerical simulation}
Three-dimensional numerical simulation using COMSOL\texttrademark \hspace{.1cm} software enabled a comparison with the experimental data and the study of parameter configurations not achieved experimentally. This entailed solving time-dependent Navier-Stokes equations under the incompressibility condition.

\begin{figure}[h]
\centering
\resizebox{.9\textwidth}{!}{%
  \includegraphics{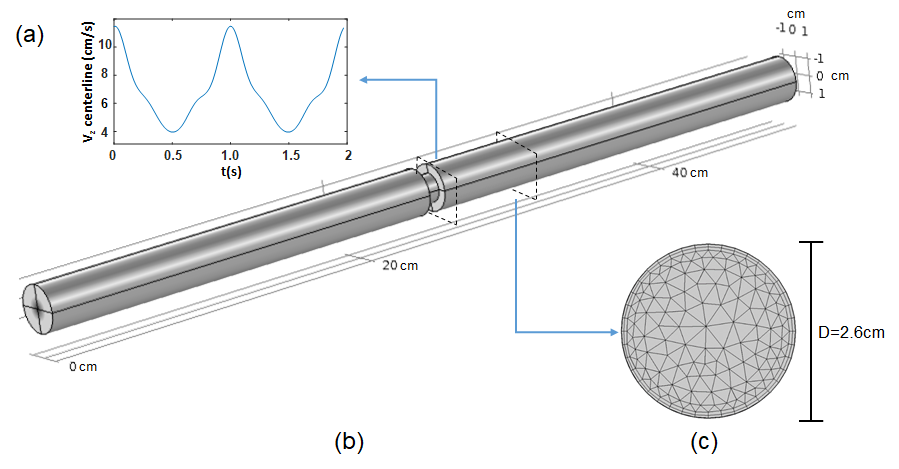}}
\caption{For $\Bar{Re}$ = 1002: (a) axial velocity at $r$ = 0 and $z$ = 0; (b) schematic drawing of the flow development section and the axisymmetric constricted tube; (c) cross sectional view of the mesh.}
\label{fig:malla}       % Give a unique label
\end{figure}

The inlet condition was defined as normal inflow and taken from experimental data in the upstream region, where the flow is developed. In order to do so, a fourth-order Fourier decomposition of the experimental flow rate was carried out. Higher-order terms were neglected, as they do not contribute significantly to the shape of velocity profiles. In order to improve numerical stability, the velocity profile calculated by decomposition was multiplied by a ramp function of time increasing steadily from 0 to 1 in 0.6 seconds. Figure \ref{fig:malla}(a) shows the axial velocity at the centreline at the  downstream edge of the constriction and presents a good agreement with its experimental analogue shown in Fig. \ref{fig:setup} (c). Outlet conditions were set to $p = 0$, being $p$ the difference between the outlet pressure and atmospheric pressure, disabling the normal flow and backflow suppression settings. No slip conditions were imposed on the inner surface of the tube wall or the constriction. A scheme of the geometry of the constricted tube used in the numerical analysis is shown in Fig. \ref{fig:malla}(b). The simulations were initialized with the fluid at rest and were run for 20 periods. The first 4 periods were discarded in order to neglect transient effects. The step size for the data output was selected to be the same of the experiments.

The mesh used was a free tetrahedral mesh composed of 351569 elements in total, with an average element quality of 0.6244 and an element volume ratio of $6.358\times 10^{-4}$. It was built using a COMSOL predefined physics-controlled mesh, enabling the normal element size settings so as to allow the mesh to adapt to the physics at specific regions. The mesh was found to be adequate by comparing results to those obtained in a mesh of 951899 elements for the inlet conditions given by $Re_\mathrm{u}$=820 and $Re_\mathrm{u}$=1187. A cross-sectional view of the tube and the mesh elements is shown in Fig. \ref{fig:malla} (c).

\section{Results and Discussion}

\subsection{Flow Pattern}
The velocity at the centreline, $r$ = 0, at the downstream edge of the constriction, $z$ = 0, as a function of time (see Fig. \ref{fig:setup}(c)) was used to establish the time reference for all the experiments. The decelerating phase (diastolic phase) was defined as $0 < t < 0.5 T$ and the accelerating phase (systolic phase) as $0.5 T < t < T$.

\begin{figure}[h!]
\resizebox{0.85\textwidth}{!}{%
\includegraphics{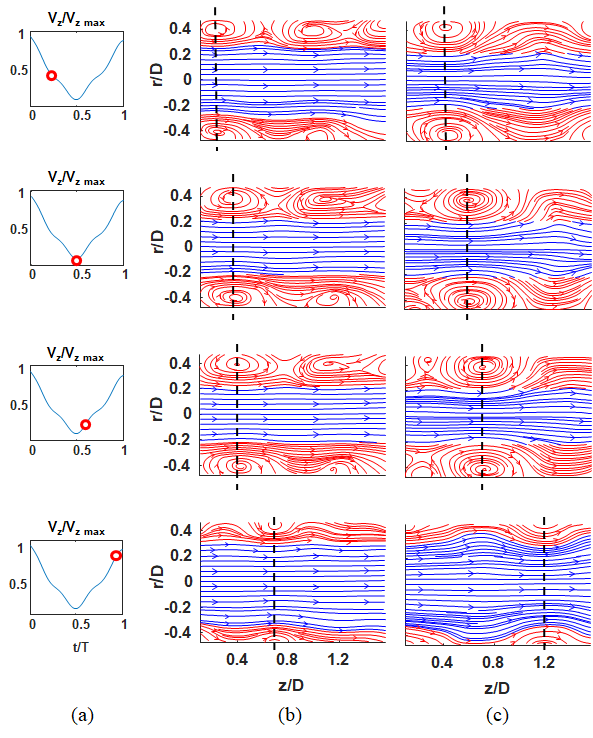}}
\centering
\caption{%{\color{blue}FALTA ARREGLAR VZ (adimensionalizar y etiqueta)} 
 (a) Axial velocity profile at $z$ = 0 and $r$ = 0 and time at which the velocity field was obtained (red circles); (b) streamlines derived from experimental data for $\Bar{Re}$ = 654, with $d_0$ = 1.6 cm; (c) streamlines derived from experimental data for $\Bar{Re}$ = 1002, with $d_0$ = 1.6 cm. In both cases, red streamlines represent the recirculation zone and blue streamlines the central jet (colours available online only).}
\label{fig:muchas_stream1}       % Give a unique label
\end{figure}
%\end{landscape}

\begin{figure}[h!]
\resizebox{0.85\textwidth}{!}{%
\includegraphics{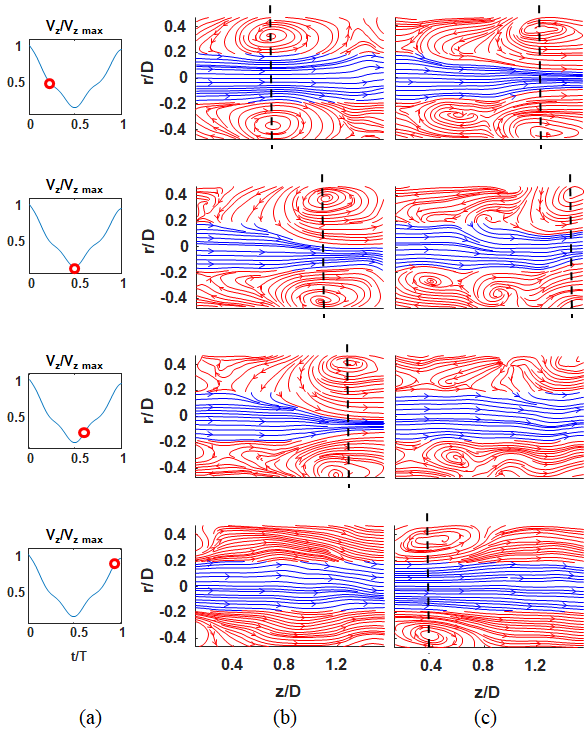}}
\centering
\caption{%{\color{blue}FALTA ARREGLAR VZ (adimensionalizar y etiqueta)}
(a) Axial velocity profile at z = 0 and r = 0 and time at which the velocity field was obtained (red circles); (b) streamlines derived from experimental data for $\Bar{Re}$ = 1106, with $d_0$ = 1.0 cm; (c) streamlines derived from experimental data for $\Bar{Re}$ = 1767, with $d_0$ = 1.0 cm. In both cases, red streamlines represent the recirculation zone and blue streamlines the central jet (colours available online only).}
\label{fig:muchas_stream2}       % Give a unique label
\end{figure}

\begin{figure}[h!]
\resizebox{0.85\textwidth}{!}{%
\includegraphics{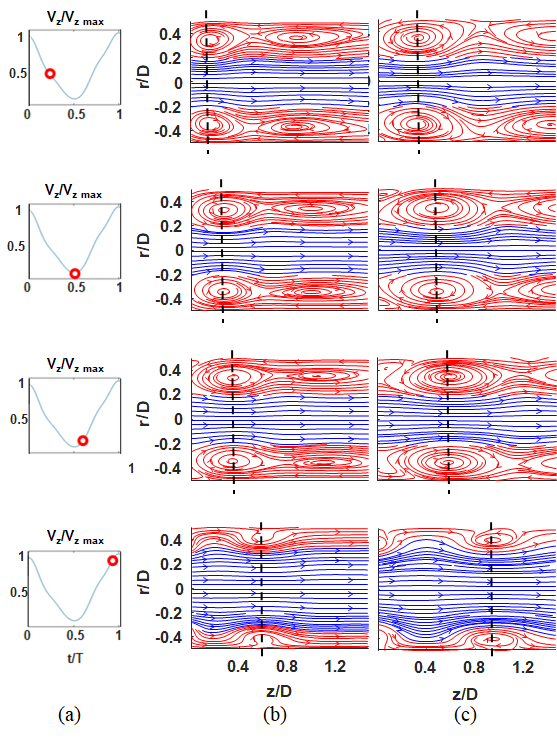}}
\centering
\caption{(a) Axial velocity profile at $z$ = 0 and $r$ = 0 and time at which the velocity field was obtained (red circles); (b) streamlines derived from simulation results for $\Bar{Re}$ = 654; (c) streamlines derived from simulation results for $\Bar{Re}$ = 1002. In both cases, red streamlines represent the central jet layer and blue streamlines represent the central jet (colours available online only).}
\label{fig:muchas_stream_sim}       % Give a unique label
\end{figure}

Figures \ref{fig:muchas_stream1} and \ref{fig:muchas_stream2} show the temporal flow patterns obtained from the experimental data corresponding to $\Bar{Re}$=654, 1002, 1106, 1767 and 2044, while Fig. \ref{fig:muchas_stream_sim} shows the patterns obtained via simulation. Figures \ref{fig:muchas_stream1}(a), \ref{fig:muchas_stream2}(a) and \ref{fig:muchas_stream_sim}(a) show the flow evolution at times 0.25 $T$, 0.5 $T$, 0.6 $T$ and 0.9 $T$. Some features were common to all the experiments and numerical results. During the accelerating phase, two main flow structures can be distinguished: a central, high-velocity central jet and a recirculation zone between the former and the inner surface of the tube wall, with vortices developing in the latter. In Figs. \ref{fig:muchas_stream1}, \ref{fig:muchas_stream2} and \ref{fig:muchas_stream_sim}, the central jet is shown in blue and the recirculation zone in red. The central jet is distinguished from recirculation zone for having vorticity values below the threshold of 30\% of the maximum of the absolute value of vorticity in each timestep. The aim of this criteria, is to approximately separate regions rather than giving a precise location of such regions. Within the recirculation zone, vortices separated from the wall and travelled along the tube. %During the accelerating phase, vortices travelled within a layer of thickness approximately equal to the step height, $h$, with $h = (D - d_0)/2$. 

Other features are dependent on the Reynolds number. For instance, for $\Bar{Re}$ = 654 (Fig. \ref{fig:muchas_stream1} (b)) the vortex had already developed at the first half of the decelerating phase and propagated until the beginning of the accelerating phase, as suggested by the dashed lines. At this stage, the thickness of the recirculation zone is approximately equal to $h$. During the accelerating phase, the thickness of the recirculation zone began to decrease as that of the central jet increased. Toward the end of this phase, the central jet was thicker and the previous vortex had nearly entirely dissipated, while a new vortex had begun to develop in the vicinity of the constriction ($z/D < 0.2$).

For $\Bar{Re}$ = 1002(Fig. \ref{fig:muchas_stream1} (c)), the flow presented the same characteristics as for $\Bar{Re}$ = 654. Here, the vortex travelled faster during the whole cycle and the recirculation zone was 27\% thicker. This is ascribed to the increment in the peak velocity of the central jet and the consequent increase in shear stress, leading to an enlargement of the recirculation zone and an increase in radial velocity. 

At $\Bar{Re}$ = 654 and $\Bar{Re}$ = 1002 two vortices were observed in the region of interest over one pulsatile period, Figs. \ref{fig:muchas_stream1}(b) and (c). In the case of $\Bar{Re}$ = 1106 (Fig. \ref{fig:muchas_stream2} (b)), only one vortex was observed over one period.  This difference arise because vortex propagates at lower velocities for lower Reynolds values, then is expected to observe two vortex in the region of interest which were shed in consecutive pulsatile periods. For $\Bar{Re}$ = 1106 at the beginning of the accelerating phase, the central jet decreased in thickness and the recirculation zone has become enlarged to occupy nearly the entire tube section at $z/D \geq 1$. This can be attributed to a marked deceleration of the central jet at the start of the accelerating phase. Due to mass conservation, the radial velocity is expected to increase, leading to the enlargement of the recirculation zone. This is also consistent with that reported previously by Sherwin \cite{shadden_2005}.

At $\Bar{Re}$ = 1767 (Fig. \ref{fig:muchas_stream2} (c)), the vortex developed and propagated faster than in the previous cases, and a secondary vortex is observed. In the decelerating phase, the vortex reached its maximum size and almost escaped from the region of interest. When the velocity of the central jet at the constriction was near its minimum, the vortex left the region of interest and the rest of the flow became disordered. In the accelerating phase, the vortex developed earlier than at lower Reynolds numbers.

A comparison of Figs. \ref{fig:muchas_stream1} and \ref{fig:muchas_stream2} illustrates how the flow pattern changed with the degree of constriction. While in Fig. \ref{fig:muchas_stream1} the vortex travelled without a substantial change in size, Fig. \ref{fig:muchas_stream2} (b) shows a sharp change in vortex size, measured radially, and Fig. \ref{fig:muchas_stream2} (c) shows that the vortex had nearly entirely dissipated at $t = 0.5 T$. Moreover, due to the reduction in $d_0$ (Fig. \ref{fig:muchas_stream2}) the velocity of the central jet was higher, and the recirculation zone became enlarged with increasing $z$, which explains the increase in vortex size. This is consistent with that reported by Sherwin et al. \cite{sherwin_2005} and Usmani \cite{usmani}.

The enlargement of the recirculation zone is present through the experiments. This could be explained in terms of circulation as pointed out in the work of Gharib et al. \cite{gharib} which studies the flow led by a moving piston into an unbounded domain. Such work states that vortex formed up to a limited amount of circulation before it sheds from the layer where is created. The main difference with our work consist on the confinement that the walls impose to the vortex. Giving as result that vortex which sheds with a certain amount of circulation enlarges in the axial direction. This mechanism can also explain the generation of secondary vortex. Specifically, vortex sheds after a precise amount of circulation is reached. Then, any excess of circulation generated goes to a trail of vorticity behind the vortex, eventually identified as secondary vortex.

Finally, experimental and numerical results were compared in order to validate the simulation. Figure \ref{fig:muchas_stream_sim} shows the evolution of the simulated flow at $\Bar{Re}$ = 654 and $\Bar{Re}$ = 1002. Comparison of Fig. \ref{fig:muchas_stream_sim} with Fig. \ref{fig:muchas_stream1} confirms a similar behaviour for the experimental and simulated flows. The main structures, i.e. central high-velocity jet and recirculation zone, were satisfactorily reproduced by numerical simulation. As expected, the flow pattern had a sharper definition in the simulation. Nevertheless, the relative error between the numerical and experimental velocity fields was less than 9\%, showing a satisfactory degree of agreement.

\subsection{Vortex Propagation}

Vortex propagation was studied by measuring the vortex displacement as a function of time along one pulsatile period. To this end, the study region was constrained to $0.3<r/D<0.5$ in order to isolate the vortex fraction forming in the superior wall of the tube. In this region vorticity values of the vortex are positive. For each time frame, the vorticity field was used to extract the vortex position. Then, in each frame, all vorticity values below a threshold of 30\% of the maximum vorticity value were disregarded to obtain the filtered vorticity field $\bar{\zeta}(r,z)$. Then, the vortex position was calculated as $(\bar{r},\bar{z})=\sum(r,z)\bar{\zeta}(r,z)$. The vortex propagation was along $z$ and is specified in figs. \ref{fig:muchas_stream1}, \ref{fig:muchas_stream2} and \ref{fig:muchas_stream_sim} by a black dashed line. This aforementioned method enables us to track the vortex and finally to measure the vortex maximum displacement which will be discussed on the next subsections.

\subsection{Numerical results for varying Womersley number}

The maximum vortex displacement was defined as $z^*/D$. The position $z^*$ is where vorticity becomes lower than 30\% of the maximum vorticity and is measured from $z_0$, the position where the vortex formed. The position $z_0$ was defined as the location of the vortex centre before it sheds. The dependence of the vortex displacement over its lifetime on $\alpha$, or $f$, was studied numerically for fixed values of $\Bar{Re}$ = 654 and $\Bar{Re}$ = 1002 and pulsatile frequency values of  0.5 $f$, 0.75 $f$, $f$, 1.5 $f$ and 2 $f$, where $f$ is the pulsatile frequency tested experimentally. 

\begin{figure}[h!]
\resizebox{.5\textwidth}{!}{%
  \includegraphics{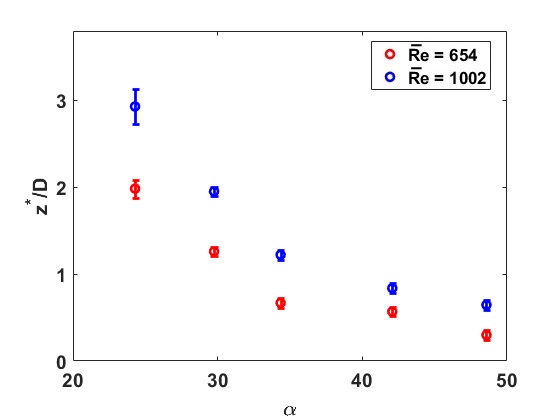}}
  \centering
\caption{Dependence of the dimensionless maximum vortex displacement on $\alpha$ (numerical results). Data for $\Bar{Re}$ = 654 (red circles) and $\Bar{Re}$ = 1002(blue squares)  }
\label{fig:desp_womersley}       % Give a unique label
\end{figure}

Figure \ref{fig:desp_womersley} clearly illustrates the dependence of the maximum vortex displacement on $\alpha$. For a fixed value of $\Bar{Re}$ , as $\alpha$ increased, i.e., $f$ increased, the flow behaviour tended to replicate within a smaller region of the tube over a shorter time span. With increasing $\alpha$, vortices tended to have a shorter lifespan and their maximum displacement was therefore smaller. In other words, $z^*/D$ decreased with increasing pulsatile frequency.

\subsection{Scaling law}

 A full description of vortex displacement was made on previous sections. Specifically, it was studied the maximum displacement of the vortex, that is the distance it travels before it vanishes. Then, we propose a scaling law which summarizes the behavior of $z^*/D$ as function of the involved parameters, $\Bar{Re}$  and $\alpha$. 

The physical dimensions for the relevant variables are $\nu$, $D$, $f$ and $v_0$.  Based on the Vaschy-Buckingham theorem, it is possible to describe $z^*/D$ as a function of two independent dimensionless numbers, $\Bar{Re}$  and $\alpha$ which involve the mentioned relevant variables. The maximum displacement $z^*/D$ was found to be proportional to the flow velocity $v_0$ (i.e. $\Bar{Re}$ ) and inversely proportional to the pulsation frequency $f$ (i.e. $\alpha^2$), suggesting that

\begin{equation}
    z^*/D=K\frac{\Bar{Re}}{\alpha^2}
    \label{eq:scale}
\end{equation}
 
where $K$ is the proportionality constant. 

\begin{figure}[h!]
\resizebox{0.5\textwidth}{!}{%
  \includegraphics{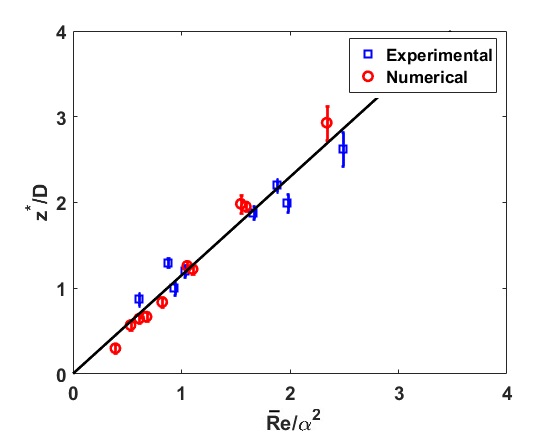}}
  \centering
\caption{Dimensionless maximum vortex displacement, $z^*/D$, as a function of the dimensionless parameter $\Bar{Re}/\alpha^2$ for experimental data (blue squares) and numerical data (red circles). %Maximum vortex displacement measured from the initial vortex position $z_0$.}%, with $z^* = z_{\mathrm{max}}-z_0$.
}
\label{fig:colapse}       % Give a unique label
\end{figure}

Figure \ref{fig:colapse} shows $z^*/D$ as a function of $\Bar{Re}/\alpha^2$ , showing that both the experimental data and the numerical results provide a good fit to Eq. 1. The fitted slope was $K = 1.15 \pm 0.19$ at a confidence level of 95\%. This result, remarks the dependence of $z^*/D$ with $\Bar{Re}$  and $1/\alpha^2$ as could be inferred from data of figure \ref{fig:desp_womersley}. 

The work of Gharib et al. \cite{gharib} shows that in the developing process of the vortex, it reaches a threshold of circulation before it sheds. In case an excess of circulation is generated, this could derive in the formation of other structures such as secondary vortex. This means that pulsatile frequency $f$ coincides with the shedding frequency of the primary vortex, the one that was tracked. Hence, the parameter  $\Bar{Re}/\alpha^2$  can be related to the Strouhal number through $\Bar{Re}/\alpha^2=\frac{2}{\pi}\frac{d_0}{D}\frac{1}{\mathrm{Sr}}$. A discussion of the results in terms of Strouhal number clarifies the relation between the oscillatory component of the flow and the stationary component of the flow. For lower values of $\Bar{Re}/\alpha^2$ , that is $\mathrm{Sr}$ close to 1, stationary and oscillatory components are comparable, which explain vortex with low to null displacement. The extreme case occurs when the flow oscillates but no vortex are shed. For instance, this could be attained for Reynolds below the used. As $\Bar{Re}/\alpha^2$  increases, that is for $\mathrm{Sr}\lesssim 10^{-1}$, vortex are shed and travel further before it vanishes as consequence of a stationary component of the flow greater than the oscillatory component. Parameter $\Bar{Re}/\alpha^2$  also states the kinematic nature of vortex displacement, by rewriting $\frac{\Bar{Re}}{\alpha^2}=\frac{2d_0\bar{v}}{\pi D^2f}$ we conclude that $z^*/D$ depends on kinematics variables and not on viscosity.

Finally, this shows that Eq. \ref{eq:scale} can be considered as a scaling law that describes the vortex displacement over its lifetime for any combination of the relevant parameters $\Bar{Re}$  and $\alpha$ on the range studied  for this particular shape of constriction.  Specifically for the dependence on the constriction shape, preliminary results with a Guassian-shaped constriction were obtained numerically, giving a difference below 20\% on the maximun vortex displacement for $Re_\mathrm{u}=$1187. We encourage further research regarding the dependence on constriction shape.

\section{Conclusions}
In this work, a pulsatile flow in an axisymmetric constricted geometry was studied experimentally and results were compared with those obtained numerically for the same values of $\Bar{Re}$  and $\alpha$. After a validation of numerical results, simulations were run over a range of $\alpha$ values that could not be tested experimentally in order to identify trends in the flow behaviour with varying $\alpha$. 

The flow structure was found to consist of a central jet around the centreline and a recirculation zone adjacent to the wall, with vortices shedding in the latter. The analysis addressed how the vortex trajectory and size changed with the system parameters.  Specifically, for a fixed value of $\alpha$, vortex size grows while decreasing $d_0$, and vortex displacement was larger for increasing $\Bar{Re}$  values. The dependence with $\alpha$, stated that as $\alpha$ increased, the behavior of the flow was reproduced in a shorter extension of the tube, that means for lower $z/D$ values. The vortex trajectory was tracked and the vortex displacement over its lifetime determined. Results showed that the vortex displacement over its lifetime decreased with increasing $\alpha$. The analysis led to a scaling law establishing a linear dependence of the vortex displacement on a dimensionless parameter combining $\Bar{Re}$  and $\alpha$, namely $\Bar{Re}/\alpha^2$ . Moreover, this parameter was found to be proportional to the inverse of Strouhal number. This relates directly the vortex behavior with the ratio between the pulsatile component and the stationary component of the flow.

As seen from the medical perspective on the issue of stenosed arteries, this results provide insight of the vortex shedding and displacement on a simplified model of stenosed arteries. This becomes crucial since vortex shedding precedes turbulence  and turbulence is related to plaque complications finally leading to cardiovascular stroke. The authors encourage further research into the behaviour of vortices over a wider range of parameters, including different constriction shapes and sizes.

\section{Acknowledgements}
This research was supported by CSIC I+D Uruguay, through the I+D project 2016 ``Estudio din\'amico de un flujo puls\'atil y sus implicaciones hemodin\'amicas vasculares", ANII (Doctoral scholarship POSNAC-2015-1-109843), ECOS-SUD (project reference number U14S04) and PEDECIBA, Uruguay. The manuscript was edited  by Eduardo Speranza.

%\bibliographystyle{apsrev4-1}
% \bibliography{biblio_paper_tubo_1.bib}

\begin{thebibliography}{33}%
	\makeatletter
	\providecommand \@ifxundefined [1]{%
		\@ifx{#1\undefined}
	}%
	\providecommand \@ifnum [1]{%
		\ifnum #1\expandafter \@firstoftwo
		\else \expandafter \@secondoftwo
		\fi
	}%
	\providecommand \@ifx [1]{%
		\ifx #1\expandafter \@firstoftwo
		\else \expandafter \@secondoftwo
		\fi
	}%
	\providecommand \natexlab [1]{#1}%
	\providecommand \enquote  [1]{``#1''}%
	\providecommand \bibnamefont  [1]{#1}%
	\providecommand \bibfnamefont [1]{#1}%
	\providecommand \citenamefont [1]{#1}%
	\providecommand \href@noop [0]{\@secondoftwo}%
	\providecommand \href [0]{\begingroup \@sanitize@url \@href}%
	\providecommand \@href[1]{\@@startlink{#1}\@@href}%
	\providecommand \@@href[1]{\endgroup#1\@@endlink}%
	\providecommand \@sanitize@url [0]{\catcode `\\12\catcode `\$12\catcode
		`\&12\catcode `\#12\catcode `\^12\catcode `\_12\catcode `\%12\relax}%
	\providecommand \@@startlink[1]{}%
	\providecommand \@@endlink[0]{}%
	\providecommand \url  [0]{\begingroup\@sanitize@url \@url }%
	\providecommand \@url [1]{\endgroup\@href {#1}{\urlprefix }}%
	\providecommand \urlprefix  [0]{URL }%
	\providecommand \Eprint [0]{\href }%
	\providecommand \doibase [0]{http://dx.doi.org/}%
	\providecommand \selectlanguage [0]{\@gobble}%
	\providecommand \bibinfo  [0]{\@secondoftwo}%
	\providecommand \bibfield  [0]{\@secondoftwo}%
	\providecommand \translation [1]{[#1]}%
	\providecommand \BibitemOpen [0]{}%
	\providecommand \bibitemStop [0]{}%
	\providecommand \bibitemNoStop [0]{.\EOS\space}%
	\providecommand \EOS [0]{\spacefactor3000\relax}%
	\providecommand \BibitemShut  [1]{\csname bibitem#1\endcsname}%
	\let\auto@bib@innerbib\@empty
	%</preamble>
	\bibitem [{\citenamefont {Naghavi}\ \emph {et~al.}(2003)\citenamefont {Naghavi}
		\emph {et~al.}}]{naghavi}%
	\BibitemOpen
	\bibfield  {author} {\bibinfo {author} {\bibfnamefont {M.}~\bibnamefont
			{Naghavi}} \emph {et~al.},\ }\bibfield  {title} {\enquote {\bibinfo {title}
			{From vulnerable plaque to vulnerable patients. a call for new definitions
				and risk assesment strategies part 1},}\ }\href@noop {} {\bibfield  {journal}
		{\bibinfo  {journal} {Circulation}\ }\textbf {\bibinfo {volume} {108}},\
		\bibinfo {pages} {1664--1772} (\bibinfo {year} {2003})}\BibitemShut {NoStop}%
	\bibitem [{\citenamefont {Muller}\ \emph {et~al.}(1989)\citenamefont {Muller},
		\citenamefont {Tofler},\ and\ \citenamefont {Stone}}]{muller}%
	\BibitemOpen
	\bibfield  {author} {\bibinfo {author} {\bibfnamefont {J.E.}\ \bibnamefont
			{Muller}}, \bibinfo {author} {\bibfnamefont {G.H}\ \bibnamefont {Tofler}}, \
		and\ \bibinfo {author} {\bibfnamefont {P.H.}\ \bibnamefont {Stone}},\
	}\bibfield  {title} {\enquote {\bibinfo {title} {Circadian variation and
				triggers of onset of acute cardiovascular disease},}\ }\href@noop {}
	{\bibfield  {journal} {\bibinfo  {journal} {Circulation}\ }\textbf {\bibinfo
			{volume} {79}},\ \bibinfo {pages} {733--743} (\bibinfo {year}
		{1989})}\BibitemShut {NoStop}%
	\bibitem [{\citenamefont {Loree}\ \emph {et~al.}(1991)\citenamefont {Loree},
		\citenamefont {Kamm}, \citenamefont {Atkinson},\ and\ \citenamefont
		{Lee}}]{loree}%
	\BibitemOpen
	\bibfield  {author} {\bibinfo {author} {\bibfnamefont {H.M.}\ \bibnamefont
			{Loree}}, \bibinfo {author} {\bibfnamefont {R.D.}\ \bibnamefont {Kamm}},
		\bibinfo {author} {\bibfnamefont {C.M.}\ \bibnamefont {Atkinson}}, \ and\
		\bibinfo {author} {\bibfnamefont {R.T.}\ \bibnamefont {Lee}},\ }\bibfield
	{title} {\enquote {\bibinfo {title} {Turbulent pressure fluctuations on
				surface of model vascular stenosis},}\ }\href@noop {} {\bibfield  {journal}
		{\bibinfo  {journal} {American Journal of Physiology}\ }\textbf {\bibinfo
			{volume} {261}},\ \bibinfo {pages} {644--650} (\bibinfo {year}
		{1991})}\BibitemShut {NoStop}%
	\bibitem [{\citenamefont {Gertz}\ and\ \citenamefont {Roberts}(1990)}]{gertz}%
	\BibitemOpen
	\bibfield  {author} {\bibinfo {author} {\bibfnamefont {S.~D.}\ \bibnamefont
			{Gertz}}\ and\ \bibinfo {author} {\bibfnamefont {W.C.}\ \bibnamefont
			{Roberts}},\ }\bibfield  {title} {\enquote {\bibinfo {title} {Hemodynamic
				shear force in rupture of coronary arterial athersoclerosis plaques},}\
	}\href@noop {} {\bibfield  {journal} {\bibinfo  {journal} {American Journal
				of Cardiology}\ }\textbf {\bibinfo {volume} {66}},\ \bibinfo {pages}
		{1368--1372} (\bibinfo {year} {1990})}\BibitemShut {NoStop}%
	\bibitem [{\citenamefont {Barger}\ \emph {et~al.}(1984)\citenamefont {Barger},
		\citenamefont {3rd}, \citenamefont {Lainey},\ and\ \citenamefont
		{Silverman}}]{barger}%
	\BibitemOpen
	\bibfield  {author} {\bibinfo {author} {\bibfnamefont {A.C.}\ \bibnamefont
			{Barger}}, \bibinfo {author} {\bibfnamefont {R.~Beeuwkes}\ \bibnamefont
			{3rd}}, \bibinfo {author} {\bibfnamefont {L.L.}\ \bibnamefont {Lainey}}, \
		and\ \bibinfo {author} {\bibfnamefont {K.~J.}\ \bibnamefont {Silverman}},\
	}\bibfield  {title} {\enquote {\bibinfo {title} {Hypothesis: vasa vasorum and
				neovascularization of human coronary arteries. a possible role in the
				pathophysiology of atheriosclerosis},}\ }\href@noop {} {\bibfield  {journal}
		{\bibinfo  {journal} {N. Engl. J. Med}\ }\textbf {\bibinfo {volume} {19}},\
		\bibinfo {pages} {175--177} (\bibinfo {year} {1984})}\BibitemShut {NoStop}%
	\bibitem [{\citenamefont {Falk}\ \emph {et~al.}(1995)\citenamefont {Falk},
		\citenamefont {Shah},\ and\ \citenamefont {Fuster}}]{falk}%
	\BibitemOpen
	\bibfield  {author} {\bibinfo {author} {\bibfnamefont {E.}~\bibnamefont
			{Falk}}, \bibinfo {author} {\bibfnamefont {P.K.}\ \bibnamefont {Shah}}, \
		and\ \bibinfo {author} {\bibfnamefont {V.}~\bibnamefont {Fuster}},\
	}\bibfield  {title} {\enquote {\bibinfo {title} {Coronary plaque
				disruption},}\ }\href@noop {} {\bibfield  {journal} {\bibinfo  {journal}
			{Circulation}\ }\textbf {\bibinfo {volume} {92}},\ \bibinfo {pages}
		{657--371} (\bibinfo {year} {1995})}\BibitemShut {NoStop}%
	\bibitem [{\citenamefont {Imoto}\ \emph {et~al.}(2005)\citenamefont {Imoto}
		\emph {et~al.}}]{imoto}%
	\BibitemOpen
	\bibfield  {author} {\bibinfo {author} {\bibfnamefont {K.}~\bibnamefont
			{Imoto}} \emph {et~al.},\ }\bibfield  {title} {\enquote {\bibinfo {title}
			{Longitudinal structural determinants of atherosclerotic plaque
				vulnerability: a computational analysis of stress distribution using vessel
				models and three-dimensional intravascular ulrasound imaging},}\ }\href@noop
	{} {\bibfield  {journal} {\bibinfo  {journal} {Journal of the American
				College of Cardiology}\ }\textbf {\bibinfo {volume} {18}},\ \bibinfo {pages}
		{1507--1515} (\bibinfo {year} {2005})}\BibitemShut {NoStop}%
	\bibitem [{\citenamefont {Caro}\ \emph {et~al.}(1971)\citenamefont {Caro} \emph
		{et~al.}}]{caro}%
	\BibitemOpen
	\bibfield  {author} {\bibinfo {author} {\bibfnamefont {C.G.}\ \bibnamefont
			{Caro}} \emph {et~al.},\ }\bibfield  {title} {\enquote {\bibinfo {title}
			{Atheroma and arterial wall shear observations, correlation and proposal of a
				shear dependent mass transfer mechanism for atherogenesis},}\ }\href@noop {}
	{\bibfield  {journal} {\bibinfo  {journal} {Proc. R. Soc. London Ser B}\
		}\textbf {\bibinfo {volume} {17}},\ \bibinfo {pages} {109--159} (\bibinfo
		{year} {1971})}\BibitemShut {NoStop}%
	\bibitem [{\citenamefont {Ku}\ \emph {et~al.}(1985)\citenamefont {Ku} \emph
		{et~al.}}]{ku_1985}%
	\BibitemOpen
	\bibfield  {author} {\bibinfo {author} {\bibfnamefont {D.~N.}\ \bibnamefont
			{Ku}} \emph {et~al.},\ }\bibfield  {title} {\enquote {\bibinfo {title}
			{Pulsatile flow and atherosclerosis in the human carotid bifurcation},}\
	}\href@noop {} {\bibfield  {journal} {\bibinfo  {journal} {Arteriosclerosis}\
		}\textbf {\bibinfo {volume} {5}},\ \bibinfo {pages} {293--302} (\bibinfo
		{year} {1985})}\BibitemShut {NoStop}%
	\bibitem [{\citenamefont {Nerem}\ and\ \citenamefont
		{Levesque}(1987)}]{nerem_87}%
	\BibitemOpen
	\bibfield  {author} {\bibinfo {author} {\bibfnamefont {R.~M.}\ \bibnamefont
			{Nerem}}\ and\ \bibinfo {author} {\bibfnamefont {M.J.}\ \bibnamefont
			{Levesque}},\ }\enquote {\bibinfo {title} {Fluid mechanics in
			atherosclerosis},}\ \ (\bibinfo  {publisher} {McGraw-Hill},\ \bibinfo {year}
	{1987})\ Chap.~\bibinfo {chapter} {21}, pp.\ \bibinfo {pages}
	{21.1--21.22}\BibitemShut {NoStop}%
	\bibitem [{\citenamefont {Gopalakrishnan}\ \emph {et~al.}(2014)\citenamefont
		{Gopalakrishnan}, \citenamefont {Pier},\ and\ \citenamefont
		{Biesheuvel}}]{Gopalakrishnan2014}%
	\BibitemOpen
	\bibfield  {author} {\bibinfo {author} {\bibfnamefont {Shyam~Sunder}\
			\bibnamefont {Gopalakrishnan}}, \bibinfo {author} {\bibfnamefont
			{Beno{\^{\i}}t}\ \bibnamefont {Pier}}, \ and\ \bibinfo {author}
		{\bibfnamefont {Arie}\ \bibnamefont {Biesheuvel}},\ }\bibfield  {title}
	{\enquote {\bibinfo {title} {Dynamics of pulsatile flow through model
				abdominal aortic aneurysms},}\ }\href {\doibase 10.1017/jfm.2014.535}
	{\bibfield  {journal} {\bibinfo  {journal} {Journal of Fluid Mechanics}\
		}\textbf {\bibinfo {volume} {758}},\ \bibinfo {pages} {150--179} (\bibinfo
		{year} {2014})}\BibitemShut {NoStop}%
	\bibitem [{\citenamefont {Arzani}\ \emph {et~al.}(2012)\citenamefont {Arzani}
		\emph {et~al.}}]{arzani}%
	\BibitemOpen
	\bibfield  {author} {\bibinfo {author} {\bibfnamefont {A.}~\bibnamefont
			{Arzani}} \emph {et~al.},\ }\bibfield  {title} {\enquote {\bibinfo {title}
			{In vivo validation of numerical prediction for turbulence intensity in
				aortic coarctation},}\ }\href@noop {} {\bibfield  {journal} {\bibinfo
			{journal} {Ann Biomed Eng}\ }\textbf {\bibinfo {volume} {40(4)}},\ \bibinfo
		{pages} {860--870} (\bibinfo {year} {2012})}\BibitemShut {NoStop}%
	\bibitem [{\citenamefont {Ford}\ \emph {et~al.}(2005)\citenamefont {Ford} \emph
		{et~al.}}]{ford}%
	\BibitemOpen
	\bibfield  {author} {\bibinfo {author} {\bibfnamefont {M.D.}\ \bibnamefont
			{Ford}} \emph {et~al.},\ }\bibfield  {title} {\enquote {\bibinfo {title}
			{Virtual angiography for visualization and validation of computational models
				of aneurysm hemodynamics},}\ }\href@noop {} {\bibfield  {journal} {\bibinfo
			{journal} {IEEE Transactions on Medical Imaging}\ }\textbf {\bibinfo {volume}
			{24}},\ \bibinfo {pages} {1586--1592} (\bibinfo {year} {2005})}\BibitemShut
	{NoStop}%
	\bibitem [{\citenamefont {Boussel}\ \emph {et~al.}(2009)\citenamefont {Boussel}
		\emph {et~al.}}]{boussel}%
	\BibitemOpen
	\bibfield  {author} {\bibinfo {author} {\bibfnamefont {L.}~\bibnamefont
			{Boussel}} \emph {et~al.},\ }\bibfield  {title} {\enquote {\bibinfo {title}
			{Phase-contrast magnetic resonance imaging measurements in intracranial
				aneurysms in vivo of flow patterns, velocity ffields and wall shear stress:
				comparisson with computational fluid dynamics},}\ }\href@noop {} {\bibfield
		{journal} {\bibinfo  {journal} {Magn. Reson. Med}\ }\textbf {\bibinfo
			{volume} {61}},\ \bibinfo {pages} {409--417} (\bibinfo {year}
		{2009})}\BibitemShut {NoStop}%
	\bibitem [{\citenamefont {Long}\ \emph {et~al.}(2001)\citenamefont {Long} \emph
		{et~al.}}]{long_2001}%
	\BibitemOpen
	\bibfield  {author} {\bibinfo {author} {\bibfnamefont {Q.}~\bibnamefont
			{Long}} \emph {et~al.},\ }\bibfield  {title} {\enquote {\bibinfo {title}
			{Numerical investigation of physiologically realistic pulsatile flow through
				arterial stenosis},}\ }\href@noop {} {\bibfield  {journal} {\bibinfo
			{journal} {J.Biomechanics}\ }\textbf {\bibinfo {volume} {34}},\ \bibinfo
		{pages} {1229--1242} (\bibinfo {year} {2001})}\BibitemShut {NoStop}%
	\bibitem [{\citenamefont {Stettler}\ and\ \citenamefont
		{Hussain}(1986)}]{sadan}%
	\BibitemOpen
	\bibfield  {author} {\bibinfo {author} {\bibfnamefont {J.C.}\ \bibnamefont
			{Stettler}}\ and\ \bibinfo {author} {\bibfnamefont {A.K.M.~Fazle}\
			\bibnamefont {Hussain}},\ }\bibfield  {title} {\enquote {\bibinfo {title} {On
				transition of pulsatile pipe flow},}\ }\href@noop {} {\bibfield  {journal}
		{\bibinfo  {journal} {J. Fluid Mech.}\ }\textbf {\bibinfo {volume} {170}},\
		\bibinfo {pages} {169--197} (\bibinfo {year} {1986})}\BibitemShut {NoStop}%
	\bibitem [{\citenamefont {Peacock}\ \emph {et~al.}(1998)\citenamefont
		{Peacock}, \citenamefont {Jones},\ and\ \citenamefont {Lutz}}]{peacock}%
	\BibitemOpen
	\bibfield  {author} {\bibinfo {author} {\bibfnamefont {J.}~\bibnamefont
			{Peacock}}, \bibinfo {author} {\bibfnamefont {T.}~\bibnamefont {Jones}}, \
		and\ \bibinfo {author} {\bibfnamefont {R.}~\bibnamefont {Lutz}},\ }\bibfield
	{title} {\enquote {\bibinfo {title} {The onset of turbulence in physiological
				pulsatile flow in a straight tube},}\ }\href@noop {} {\bibfield  {journal}
		{\bibinfo  {journal} {Experiments in Fluids}\ }\textbf {\bibinfo {volume}
			{24}},\ \bibinfo {pages} {1--9} (\bibinfo {year} {1998})}\BibitemShut
	{NoStop}%
	\bibitem [{\citenamefont {Chua}\ \emph {et~al.}(2009)\citenamefont {Chua} \emph
		{et~al.}}]{chua_piv_2009}%
	\BibitemOpen
	\bibfield  {author} {\bibinfo {author} {\bibfnamefont {C.S.}\ \bibnamefont
			{Chua}} \emph {et~al.},\ }\bibfield  {title} {\enquote {\bibinfo {title}
			{Particle image of non axisymmetic stenosis model},}\ }\href@noop {}
	{\bibfield  {journal} {\bibinfo  {journal} {8th international symposium on
				particle image velocimetry}\ } (\bibinfo {year} {2009})}\BibitemShut
	{NoStop}%
	\bibitem [{\citenamefont {Ahmed}\ and\ \citenamefont
		{Giddens}(1983)}]{ahmed_83}%
	\BibitemOpen
	\bibfield  {author} {\bibinfo {author} {\bibfnamefont {S.A.}\ \bibnamefont
			{Ahmed}}\ and\ \bibinfo {author} {\bibfnamefont {D.P.}\ \bibnamefont
			{Giddens}},\ }\bibfield  {title} {\enquote {\bibinfo {title} {Flow
				disturbance measurements through a constricted tube at moderate reynolds
				numbers},}\ }\href@noop {} {\bibfield  {journal} {\bibinfo  {journal}
			{J.Biomechanics}\ }\textbf {\bibinfo {volume} {16}},\ \bibinfo {pages}
		{955--963} (\bibinfo {year} {1983})}\BibitemShut {NoStop}%
	\bibitem [{\citenamefont {Ahmed}\ and\ \citenamefont
		{Giddens}(1984)}]{ahmed_84}%
	\BibitemOpen
	\bibfield  {author} {\bibinfo {author} {\bibfnamefont {S.A.}\ \bibnamefont
			{Ahmed}}\ and\ \bibinfo {author} {\bibfnamefont {D.P.}\ \bibnamefont
			{Giddens}},\ }\bibfield  {title} {\enquote {\bibinfo {title} {Pulsatile
				poststenotic flow studies with laser doppler anemometry},}\ }\href@noop {}
	{\bibfield  {journal} {\bibinfo  {journal} {J.Biomechanics}\ }\textbf
		{\bibinfo {volume} {17}},\ \bibinfo {pages} {695--705} (\bibinfo {year}
		{1984})}\BibitemShut {NoStop}%
	\bibitem [{\citenamefont {Isler}\ \emph {et~al.}(2018)\citenamefont {Isler},
		\citenamefont {Gioria},\ and\ \citenamefont {Carmo}}]{Isler2018}%
	\BibitemOpen
	\bibfield  {author} {\bibinfo {author} {\bibfnamefont {Jo{\~{a}}o~A.}\
			\bibnamefont {Isler}}, \bibinfo {author} {\bibfnamefont {Rafael~S.}\
			\bibnamefont {Gioria}}, \ and\ \bibinfo {author} {\bibfnamefont {Bruno~S.}\
			\bibnamefont {Carmo}},\ }\bibfield  {title} {\enquote {\bibinfo {title}
			{Bifurcations and convective instabilities of steady flows in a constricted
				channel},}\ }\href {\doibase 10.1017/jfm.2018.410} {\bibfield  {journal}
		{\bibinfo  {journal} {Journal of Fluid Mechanics}\ }\textbf {\bibinfo
			{volume} {849}},\ \bibinfo {pages} {777--804} (\bibinfo {year}
		{2018})}\BibitemShut {NoStop}%
	\bibitem [{\citenamefont {Mittal}\ \emph {et~al.}(2003)\citenamefont {Mittal}
		\emph {et~al.}}]{mittal}%
	\BibitemOpen
	\bibfield  {author} {\bibinfo {author} {\bibfnamefont {R.}~\bibnamefont
			{Mittal}} \emph {et~al.},\ }\bibfield  {title} {\enquote {\bibinfo {title}
			{Numerical study of pulsatile flow in a constricted channel},}\ }\href@noop
	{} {\bibfield  {journal} {\bibinfo  {journal} {J.Fluid Mech.}\ }\textbf
		{\bibinfo {volume} {485}},\ \bibinfo {pages} {337--378} (\bibinfo {year}
		{2003})}\BibitemShut {NoStop}%
	\bibitem [{\citenamefont {Sherwin}\ and\ \citenamefont
		{Blackburn}(2005)}]{sherwin_2005}%
	\BibitemOpen
	\bibfield  {author} {\bibinfo {author} {\bibfnamefont {S.J.}\ \bibnamefont
			{Sherwin}}\ and\ \bibinfo {author} {\bibfnamefont {H.M.}\ \bibnamefont
			{Blackburn}},\ }\bibfield  {title} {\enquote {\bibinfo {title} {Three
				dimensional instabilities and transition of steady and pulsatile axisymmetric
				stenotic flows},}\ }\href@noop {} {\bibfield  {journal} {\bibinfo  {journal}
			{J. Fluid Mech.}\ }\textbf {\bibinfo {volume} {533}},\ \bibinfo {pages}
		{297--327} (\bibinfo {year} {2005})}\BibitemShut {NoStop}%
	\bibitem [{\citenamefont {Ling}\ and\ \citenamefont {Atabek}(1972)}]{ling_72}%
	\BibitemOpen
	\bibfield  {author} {\bibinfo {author} {\bibfnamefont {S.C.}\ \bibnamefont
			{Ling}}\ and\ \bibinfo {author} {\bibfnamefont {H.B.}\ \bibnamefont
			{Atabek}},\ }\bibfield  {title} {\enquote {\bibinfo {title} {A nonlinear
				analysis of pulsatile flow in arteries},}\ }\href@noop {} {\bibfield
		{journal} {\bibinfo  {journal} {J. Fluid Mech.}\ }\textbf {\bibinfo {volume}
			{55}},\ \bibinfo {pages} {493--511} (\bibinfo {year} {1972})}\BibitemShut
	{NoStop}%
	\bibitem [{\citenamefont {Griffith}\ \emph {et~al.}(2009)\citenamefont
		{Griffith} \emph {et~al.}}]{griffith}%
	\BibitemOpen
	\bibfield  {author} {\bibinfo {author} {\bibfnamefont {M.D.}\ \bibnamefont
			{Griffith}} \emph {et~al.},\ }\bibfield  {title} {\enquote {\bibinfo {title}
			{Pulsatile flow in stenotic geometries: flow behaviour and stability},}\
	}\href@noop {} {\bibfield  {journal} {\bibinfo  {journal} {J.Fluid Mech.}\
		}\textbf {\bibinfo {volume} {622}},\ \bibinfo {pages} {291--320} (\bibinfo
		{year} {2009})}\BibitemShut {NoStop}%
	\bibitem [{\citenamefont {Usmani}\ and\ \citenamefont
		{Muralidhar}(2016)}]{usmani}%
	\BibitemOpen
	\bibfield  {author} {\bibinfo {author} {\bibfnamefont {A.Y.}\ \bibnamefont
			{Usmani}}\ and\ \bibinfo {author} {\bibfnamefont {K.}~\bibnamefont
			{Muralidhar}},\ }\bibfield  {title} {\enquote {\bibinfo {title} {Pulsatile
				flow in a compliant stenosed asymmetric model},}\ }\href@noop {} {\bibfield
		{journal} {\bibinfo  {journal} {Exp. Fluids}\ }\textbf {\bibinfo {volume}
			{57(12)}},\ \bibinfo {pages} {186} (\bibinfo {year} {2016})}\BibitemShut
	{NoStop}%
	\bibitem [{\citenamefont {Westerweel}(1997)}]{piv1}%
	\BibitemOpen
	\bibfield  {author} {\bibinfo {author} {\bibfnamefont {J.}~\bibnamefont
			{Westerweel}},\ }\bibfield  {title} {\enquote {\bibinfo {title} {Fundamentals
				of digital particle velocimetry},}\ }\href@noop {} {\bibfield  {journal}
		{\bibinfo  {journal} {Measurement science and technology}\ }\textbf {\bibinfo
			{volume} {8}},\ \bibinfo {pages} {1379--1392} (\bibinfo {year}
		{1997})}\BibitemShut {NoStop}%
	\bibitem [{\citenamefont {Raffel}\ \emph {et~al.}(2007)\citenamefont {Raffel},
		\citenamefont {Willert}, \citenamefont {Wereley},\ and\ \citenamefont
		{Kompenhans}}]{piv2}%
	\BibitemOpen
	\bibfield  {author} {\bibinfo {author} {\bibfnamefont {M.}~\bibnamefont
			{Raffel}}, \bibinfo {author} {\bibfnamefont {C.}~\bibnamefont {Willert}},
		\bibinfo {author} {\bibfnamefont {S.}~\bibnamefont {Wereley}}, \ and\
		\bibinfo {author} {\bibfnamefont {J.}~\bibnamefont {Kompenhans}},\
	}\href@noop {} {\emph {\bibinfo {title} {Particle image velocimetry, a
				practical guide.}}},\ \bibinfo {edition} {2nd}\ ed.\ (\bibinfo  {publisher}
	{Springer},\ \bibinfo {year} {2007})\BibitemShut {NoStop}%
	\bibitem [{\citenamefont {Cassanova}\ and\ \citenamefont
		{Giddens}(1977)}]{casanova}%
	\BibitemOpen
	\bibfield  {author} {\bibinfo {author} {\bibfnamefont {R.~A.}\ \bibnamefont
			{Cassanova}}\ and\ \bibinfo {author} {\bibfnamefont {D.~P.}\ \bibnamefont
			{Giddens}},\ }\bibfield  {title} {\enquote {\bibinfo {title} {Disorder distal
				to modeled stenoses in steady and pulsatile flow},}\ }\href@noop {}
	{\bibfield  {journal} {\bibinfo  {journal} {J. Biomechanics}\ }\textbf
		{\bibinfo {volume} {11}},\ \bibinfo {pages} {441--453} (\bibinfo {year}
		{1977})}\BibitemShut {NoStop}%
	\bibitem [{\citenamefont {Balay}(2012)}]{balay}%
	\BibitemOpen
	\bibfield  {author} {\bibinfo {author} {\bibfnamefont {G.}~\bibnamefont
			{Balay}},\ }\emph {\bibinfo {title} {Elasticidad en tejidos arteriales,
			dise\~no de un coraz\'on artficial in vitro y nuevo m\'etodo ultras\'onico de
			determinaci\'on de elasticidad arterial}},\ \href@noop {} {Master's thesis},\
	\bibinfo  {school} {Universidad de la Rep\'ublica, Uruguay} (\bibinfo {year}
	{2012})\BibitemShut {NoStop}%
	\bibitem [{\citenamefont {Ku}(1997)}]{ku_1997}%
	\BibitemOpen
	\bibfield  {author} {\bibinfo {author} {\bibfnamefont {D.N.}\ \bibnamefont
			{Ku}},\ }\bibfield  {title} {\enquote {\bibinfo {title} {Blood flow in
				arteries},}\ }\href@noop {} {\bibfield  {journal} {\bibinfo  {journal} {Ann.
				Rev. Fluid Mech}\ }\textbf {\bibinfo {volume} {29}},\ \bibinfo {pages}
		{399--434} (\bibinfo {year} {1997})}\BibitemShut {NoStop}%
	\bibitem [{\citenamefont {Shadden}\ \emph {et~al.}(2005)\citenamefont
		{Shadden}, \citenamefont {Leiken},\ and\ \citenamefont
		{Marsden}}]{shadden_2005}%
	\BibitemOpen
	\bibfield  {author} {\bibinfo {author} {\bibfnamefont {S.C.}\ \bibnamefont
			{Shadden}}, \bibinfo {author} {\bibfnamefont {F.}~\bibnamefont {Leiken}}, \
		and\ \bibinfo {author} {\bibfnamefont {J.E.}\ \bibnamefont {Marsden}},\
	}\bibfield  {title} {\enquote {\bibinfo {title} {Definition and properties of
				lagrangian coherent structures from finite-time lyapunov exponents in
				two-dimendional aperiodic flows},}\ }\href@noop {} {\bibfield  {journal}
		{\bibinfo  {journal} {Physica D}\ }\textbf {\bibinfo {volume} {212}},\
		\bibinfo {pages} {271--304} (\bibinfo {year} {2005})}\BibitemShut {NoStop}%
	\bibitem [{\citenamefont {Gharib}\ \emph {et~al.}(1998)\citenamefont {Gharib},
		\citenamefont {Rambod},\ and\ \citenamefont {Shariff}}]{gharib}%
	\BibitemOpen
	\bibfield  {author} {\bibinfo {author} {\bibfnamefont {M.}~\bibnamefont
			{Gharib}}, \bibinfo {author} {\bibfnamefont {E.}~\bibnamefont {Rambod}}, \
		and\ \bibinfo {author} {\bibfnamefont {K.}~\bibnamefont {Shariff}},\
	}\bibfield  {title} {\enquote {\bibinfo {title} {A universal time scale for
				vortex ring formation},}\ }\href@noop {} {\bibfield  {journal} {\bibinfo
			{journal} {Journal of Fluid Mechanics}\ }\textbf {\bibinfo {volume} {360}},\
		\bibinfo {pages} {121--140} (\bibinfo {year} {1998})}\BibitemShut {NoStop}%



\end{thebibliography}

\end{document}